\documentclass{article}
\usepackage{authblk}
\usepackage[sorting=none,style=numeric-comp]{biblatex}
\usepackage[breaklinks]{hyperref}
\bibliography{references}
\usepackage{graphicx} 

\usepackage[a4paper,top=1.4cm,bottom=1.9cm,left=2cm,right=2cm,marginparwidth=1.75cm]{geometry}

\title{Data-based form factor corrections between the two-pion $\tau$ and $e^+e^-$ spectral functions}

\author[1]{Michel Davier}
\author[2]{Bogdan Malaescu}
\author[1]{Zhiqing Zhang}

\affil[1]{IJCLab, Universit\'e Paris-Saclay et CNRS/IN2P3, 91405, Orsay, France}
\affil[2]{ LPNHE, Sorbonne Université, Université Paris Cité, CNRS/IN2P3, 75252, Paris, France}

\begin{document}

\maketitle

\begin{abstract}
The $\tau$ spectral functions are an alternative to $e^+e^-$ cross-sections, where different measurements are not consistent, for computing the hadronic vacuum contribution to the muon magnetic anomaly $a_\mu$. This requires a  control of isospin-breaking effects which have to be corrected for. So far these corrections have been evaluated  using theoretical models. In this letter, a new approach based only on data is presented for the determination of the most critical correction relating the $e^+e^-$ and $\tau$ pion form factors. An updated evaluation of the total isospin-breaking correction is given and its impact is discussed in the context of $e^+e^-$-based $a_\mu$ predictions and of the direct measurement.
\end{abstract}

\section{Introduction}
Hadronic vacuum polarization (HVP) contributions are crucial to compute Standard Model predictions of observables such as the muon magnetic anomaly $a_\mu=(g-2)/2$ or the running of the electromagnetic coupling $\alpha (s)$, with $s$ being the energy scale squared. It was traditionally done through dispersion relations integrating over the cross-section for $e^+e^-\rightarrow {\rm hadrons}$ with known kernel functions~\cite{Aoyama:2020ynm}. More recently, valuable progress was achieved using lattice QCD calculations~\cite{Borsanyi:2020mff,Boccaletti:2024guq,RBC:2024fic,Djukanovic:2024cmq,Bazavov:2024eou}.  Hadronic $\tau$ decays into vector states provide another source of information which can be used for computing HVP contributions under isospin symmetry, as originally proposed in Ref.~\cite{Alemany:1997tn}. Several high-quality datasets were collected more than 20 years ago at LEP and also at $B$ factories, first by CLEOc, later by Belle and BABAR. The available experimental information was very recently reconsidered~\cite{Davier:2023fpl}, in view of the poor consistency among $e^+e^-$ results.

The straightforward use of $\tau$ data relies on isospin symmetry which is known to be broken by electromagnetic interactions and also in the strong interactions due to quark mass differences. Therefore the information contained in hadronic mass spectra from $\tau$ decays has to be corrected for isospin-breaking (IB) effects before they can be inserted in dispersion relations. Among all hadronic final states to be considered for the muon $g-2$ the two-pion system plays an important role as 73\% of the full HVP contribution originates from this channel, which also yields the dominant contribution to the final uncertainty on the $a_\mu$ prediction. In this letter, we concentrate on this leading two-pion contribution.

The tau spectral function in the dominant $\tau$ decay mode $\tau^-\to \pi^-\pi^0\nu_\tau$ defined as
\begin{equation}
\label{eq:SF}
v_{\pi\pi^0}(s)=\frac{m^2_\tau}{6|V_{ud}|^2}\frac{\mathcal{B}_{\pi\pi^0}}{\mathcal{B}_e}\frac{dN_{\pi\pi^0}}{N_{\pi\pi^0}ds}\times\left[\left(1-\frac{s}{m^2_\tau}\right)^2\left(1+\frac{2s}{m^2_\tau}\right)\right]^{-1}
\end{equation}
has been precisely measured by several experiments~\cite{ALEPH:2005qgp,Davier:2013sfa,OPAL:1998rrm,CLEO:1999dln,Belle:2008xpe} under very different conditions at LEP and the $B$ factories. Here $m_\tau$ is the $\tau$ lepton mass, $|V_{ud}|$ the CKM matrix element, $\mathcal{B}_{\pi\pi^0}$ and $\mathcal{B}_e$ are the branching fractions of $\tau^-\to \pi^-\pi^0\nu_\tau(\gamma)$ (final state photon radiation is implied) and of $\tau^-\to e^-\bar{\nu}_e\nu_\tau$, respectively, and $dN_{\pi\pi^0}/N_{\pi\pi^0}ds$ is the normalized invariant mass spectrum of the hadronic final state. 
The precision achieved in the experiments for the branching fractions (0.4\%) and the agreement between the different results provide a highly precise normalization of the spectral functions, even superior to that obtained in $e^+e^-$ data. There is also good agreement between the spectral function results as shown in Ref.~\cite{Davier:2013sfa}.
These measured spectral functions have been widely used (see e.g.\ Ref.~\cite{Davier:2005xq}) for a number of applications including in particular the evaluation of the lowest-order (LO) HVP $a_\mu^\mathrm{HVP,\,LO}$ and $\Delta\alpha^{(5)}_\mathrm{had}$ as originally proposed in Ref.~\cite{Alemany:1997tn}. 

\section{Isospin-breaking corrections: a new procedure}
\label{sec:IB}

Up to now, IB corrections to the two-pion spectral function from $\tau$ decays have been estimated with theoretical models~\cite{Davier:2010fmf}. Some developments are underway using a dispersive approach~\cite{cottini-kek} and lattice~\cite{bruno-kek}. The leading energy-independent correction $S_\mathrm{EW}$ is from short-distance electroweak loops, which is well under control~\cite{Sirlin:1977sv,Sirlin:1981ie,Marciano:1988vm,Braaten:1990ef,Davier:2002dy}. All other corrections are energy-dependent: 
\begin{equation}
   R_\mathrm{IB}(s)=\frac{\mathrm{FSR}(s)}{G_\mathrm{EM}(s)}\frac{\beta^3_0(s)}{\beta^3_-(s)}\left|\frac{F_0(s)}{F_-(s)}\right|^2
\end{equation}
where $F_0(s)$ and $F_-(s)$ are the neutral and charged $\rho$ form factors, respectively. The IB-corrected $\tau$ spectral function is then obtained by multiplying the expression given in Eq.~(\ref{eq:SF}) by the factor $R_\mathrm{IB}(s)/S_\mathrm{EW}$.

The effect of the charged/neutral pion mass difference in the cross-section through the pion velocity factors $\beta_{0,-}^3$ is accurately known. The long-distance radiative correction $G_\mathrm{EM}$ is model-dependent~\cite{Cirigliano:2002pv,Flores-Baez:2006yiq}, but leads to a rather small effect. Also, since $G_\mathrm{EM}$ is applied to the measured two-pion mass spectrum including additional radiation and delivers the correction to a final $\pi^+\pi^-$ spectrum without radiation, it is necessary to re-introduce final-state radiation (FSR) which is also model-dependent. It has been evaluated so far using a scalar QED calculation~\cite{jss89}, treating charged pions as point-like objects. Then the effect of $\rho-\omega$ interference, only present in $e^+e^-$, is taken from fitting experimental data with a rather model-independent functional form, inducing a fast bipolar variation around the narrow $\omega$ resonance.

By far the most worrisome IB effects appear at the level of the pion form factor itself, usually parametrized by the mass and width of the charged and neutral $\rho$ mesons. For the mass difference only some loose theoretical estimates are available, fortunately with a small IB effect. The situation on the width difference is more challenging. So far two effects have been considered and embedded in a calculation~\cite{Flores-Baez:2007vnd} of the radiative $\rho$ width $\Gamma_{\rho\rightarrow\pi\pi\gamma}$ taking into account the charged/neutral pion difference through again a $\beta^3$ factor, the radiative calculation being done with scalar QED with point-like charged mesons. The two effects lead to corrections with opposite signs, thus reducing the overall result. However the radiative estimate may be incomplete~\footnote{Recent discussions in the muon $g-2$ Theory Initiative are acknowledged.}, thus introducing a weak element in the global IB budget. The uncertainty is difficult to ascertain in the absence of more complete calculations. In these conditions one could be tempted to assign a large uncertainty to conservatively cover theoretical ignorance, which is not a satisfactory solution.

This situation prompted us to investigate a data-based approach with minimal dependence on models. We know since decades that the pion form factor measured in $e^+e^-$ and $\tau$ data in the region of interest for HVP calculations is dominated by the broad $\rho$ resonance which is well represented by the Gounaris-Sakurai version of the Breit-Wigner line shape~\cite{Gounaris:1968mw}. This functional form depends on only two parameters, the mass $m_\rho$ and width $\Gamma_\rho$ = $\Gamma_\rho (m_\rho^2)$, and the complete description involves an energy-dependent width $\Gamma_\rho(s)$. The $\rho$ resonance shape has to be complemented by contributions from higher-mass isovector states which are small in the $\rho$ region, but not negligible, due to their long tails. At the moment the best experimental information on these contributions to the pion form factor has been obtained by BABAR with a measurement and a fit up to a mass of 3\,GeV, well above the $\rho$ region~\cite{BaBar:2012bdw}. Also valuable information is available from Belle~\cite{Hayashii:2009zz} using $\tau$ data, but it is more restricted because of the kinematic limit imposed by the $\tau$ mass. Both experiments provide unique information on the higher mass states with distinct interference patterns, which enable the characterization of their parameters.
For this analysis the more complete set of BABAR resonance parameters is used, in good agreement with the Belle values.

In usual form factor fits from either $e^+e^-$ or $\tau$ data, involving a normalization constraint in the functional form, the fitted mass and width are correlated to the corresponding $a_\mu$ dispersion integral. 
Therefore such a procedure would not be suitable to obtain a $\tau$ IB correction independent from the $e^+e^-$ result. 
In order to avoid this circularity problem one has to decouple the resonance parameters from the absolute normalization of the spectral function, {\it i.e.} determine the mass and width values only through the shape of the respective spectral functions, leaving their normalization free in the fits. 
While doing so, other aspects of the shape differences observed among various datasets are carefully studied by quantifying the quality of the fits and comparing quantitatively the fitted parameters with their uncertainties.
It is the procedure we follow in this study for all the fits of $\tau$ and $e^+e^-$ experimental data. Also, since we want to compare the two form factors it is necessary to apply the energy dependence of the $G_\mathrm{EM}$ and $\beta^3$ factors before fitting their respective shape. The small $G_\mathrm{EM}$ shape correction is the only explicit model-dependence introduced in this procedure.

\section{The fit procedure}

We recall here the details of the Gounaris-Sakurai (GS) form factor parametrization~\cite{Gounaris:1968mw}:
 \begin{equation}
 \label{eq:GS}
\mathrm{BW}^\mathrm{GS}(s, m, \Gamma) = \frac{m^2 (1 + d(m) \Gamma/m) }{m^2 - s + f(s, m, \Gamma) - i m \Gamma (s, m, \Gamma)}~,
\end{equation}
with
\begin{equation}
\Gamma (s, m, \Gamma) = \Gamma  \frac{s}{m^2} \left( \frac{\beta (s) }{ \beta (m^2)} \right) ^3~,
\end{equation}
where $\beta(s) = \sqrt{\left[1 - (m_1+m_2)^2/s\right]\left[1 - (m_1-m_2)^2/s\right]}$ with $m_1=m_2=m_{\pi^\pm}$ for $e^+e^-$ and $m_1=m_{\pi^\pm}$, $m_2=m_{\pi^0}$ for $\tau$. 

The auxiliary functions used in the GS parametrization are:
\begin{equation}
d(m) = \frac{3}{\pi} \frac{m_\pi^2}{k^2(m^2)} \ln \left( \frac{m+2 k(m^2)}{2 m_\pi} \right) 
   + \frac{m}{2\pi  k(m^2)}
   - \frac{m_\pi^2  m}{\pi k^3(m^2)}~,
\end{equation}
\begin{equation}
f(s, m, \Gamma) = \frac{\Gamma  m^2}{k^3(m^2)} \left[ k^2(s) (h(s)-h(m^2)) + (m^2-s) k^2(m^2)  h'(m^2)\right]~,
\end{equation}
where 
\begin{eqnarray}
k(s)\!\!\!&=&\!\!\! \frac{1}{2} \sqrt{s}  \beta (s)~,\\ 
h(s)\!\!\!&=&\!\!\! \frac{2}{\pi}  \frac{k(s)}{\sqrt{s}}  \ln \left( \frac{\sqrt{s}+2 k(s)}{2 m_\pi} \right).
\end{eqnarray}
and $h'(s)$ is $h(s)$ derivative.

Contributions from higher-mass $\rho$ states are added in the complete fitting function, each with a complex amplitude relative to the main $\rho$ amplitude as performed in Ref.~\cite{BaBar:2012bdw} from which we keep and fix the parameters obtained from a fit of the $\pi^+\pi^-$ BABAR data from threshold to 3\,GeV. When applying the same relative contributions in $\tau$ fits in the $\rho$ region it is assumed that IB is negligible at this level as it would appear as a second-order effect.

Fits to $e^+e^-$ data should also include the effect of $\rho-\omega$ interference, needing four additional parameters, two for a complex amplitude, the mass $m_\omega$ and the width $\Gamma_\omega$ of the narrow $\omega$ resonance. At higher mass $\rho-\phi$ interference is also needed, but in practice only CMD-3 is covering this mass range with good statistics. To keep all $e^+e^-$ experiments on the same footing all the fits are performed below 0.9\,GeV and for ensuring consistency the same restriction is applied to $\tau$ data. Again, in both cases the higher-mass contributions are fixed and taken from the wide BABAR fit.

\section{Results from the fits}

Fits are performed using all the recent available two-pion $e^+e^-$ data taken in the scan mode at BINP, Novosibirsk (CMD-2~\cite{CMD-2:2003gqi,CMD-2:2005mvb,Aulchenko:2006dxz,CMD-2:2006gxt}, SND~\cite{Achasov:2006vp}, SND20~\cite{SND:2020nwa} CMD-3~\cite{CMD-3:2023alj}) or using the ISR method with KLOE~\cite{KLOE:2008fmq,KLOE:2010qei,KLOE:2012anl}, BABAR~\cite{BaBar:2012bdw}, and BES~\cite{BESIII:2015equ}. For two-pion $\tau$ data the available results are from LEP (ALEPH~\cite{Davier:2013sfa}, OPAL~\cite{OPAL:1998rrm}) and B-factories (CLEO~\cite{CLEO:1999dln}, Belle~\cite{Hayashii:2009zz}). The OPAL data has been used for a combined analysis of all four experiments, but with limited precision compared to the other experiments as shown in Figure 4 of Ref.~\cite{Davier:2013sfa} . For the present analysis the shape determination requires a better control of the spectral function. To our knowledge no OPAL publication is available with a fit of the $\rho$ line shape (no entry in PDG) and detailed information on relevant systematic uncertainties affecting the mass and width determination is not at hand. Therefore in this analysis only the more precise ALEPH, Belle, and CLEO results are used.

As announced, all the fits of $e^+e^-$ and $\tau$ data discussed in this section are performed with the Gounaris-Sakurai parametrization for the $\rho$ resonance, leaving the normalization free. The fit range covers the available data for each experiment, limited to a mass of 0.9\,GeV for the reasons explained above. The contributions of higher mass resonances are fixed using the parameters found in the wide-mass BABAR fit. Thus,  each fit involves three free parameters: the normalization~\footnote{Of course a theoretical form factor is constrained to 1 at $s=0$, but here we are dealing with the experimental form factor derived from the measurement of the cross-section and the normalized form factor function is only used as a parametrization of the shape. It is also worth noting that this fitted normalization factor is mostly constrained by the precise data on the $\rho$ peak, rather than the low-mass points.}, $m_\rho$ and $\Gamma_\rho$, for $\tau$ and $e^+e^-$. For the latter, four more free parameters are used to describe $\rho-\omega$ interference. For $\tau$ and ISR $e^+e^-$ experiments, data are available as histograms and the finite bin width is taken into account to properly treat rapid variations of the fitting resonance function.

\begin{figure}[htp] \centering
    \includegraphics[width=0.42\columnwidth]{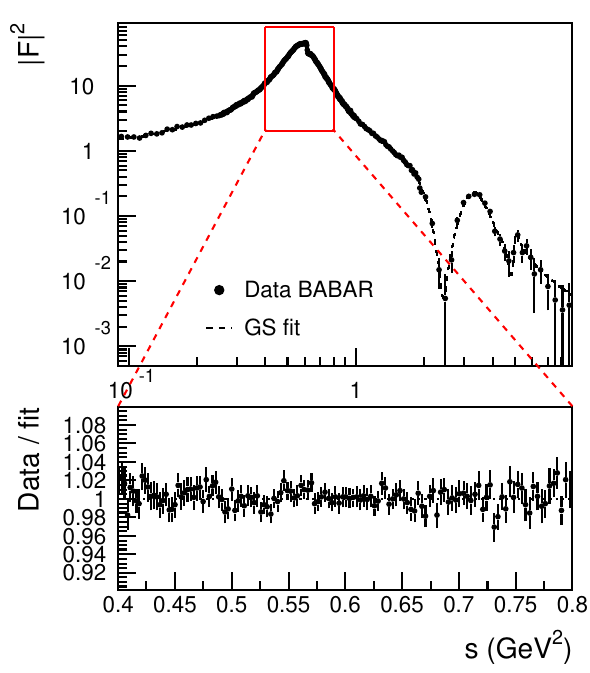}
    \includegraphics[width=0.42\columnwidth]{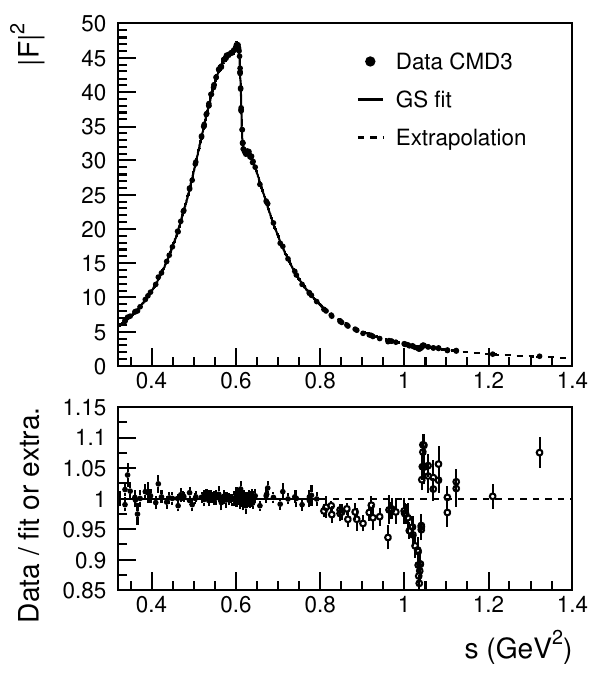}
    \includegraphics[width=0.42\columnwidth]{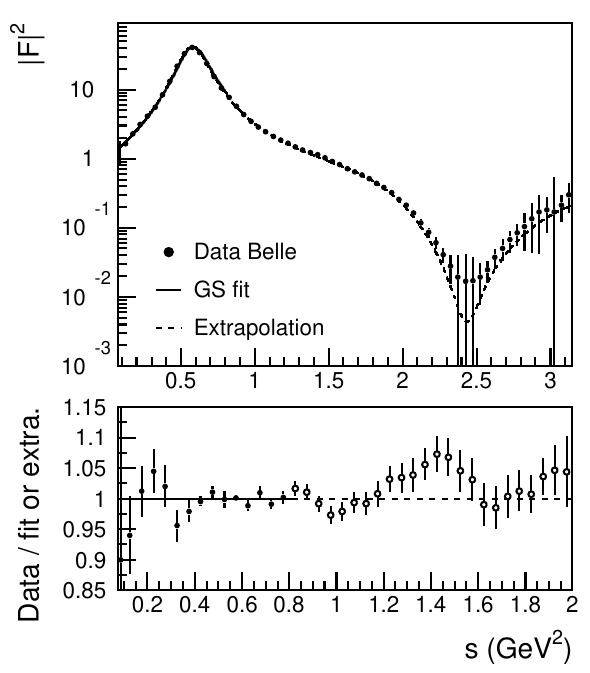}
    \caption{\label{fig:tail}   \small Top-left: a GS fit to the BABAR measurement in the full mass range with 19 free parameters. The ratio of the BABAR data over the fit in the rho resonance region is shown in the bottom panel. Top-right: a GS fit to the CMD-3 measurements below 0.9\,GeV with 7 free parameters, using the BABAR tail shown at high mass with a dashed line. The region between 0.9 and 1.05\,GeV is affected by the $\rho-\phi$ interference which is not included in our parametrization.
    Bottom: a GS fit to the Belle measurement below 0.9\,GeV with three free parameters, using the BABAR tail shown at high mass with a dashed line. For both CMD-3 and Belle, the ratio between the data and the fit allows one to check the validity of using the BABAR tail for these experiments.
}
\end{figure}

\begin{figure}[tbp] \centering
    \includegraphics[width=0.42\columnwidth]{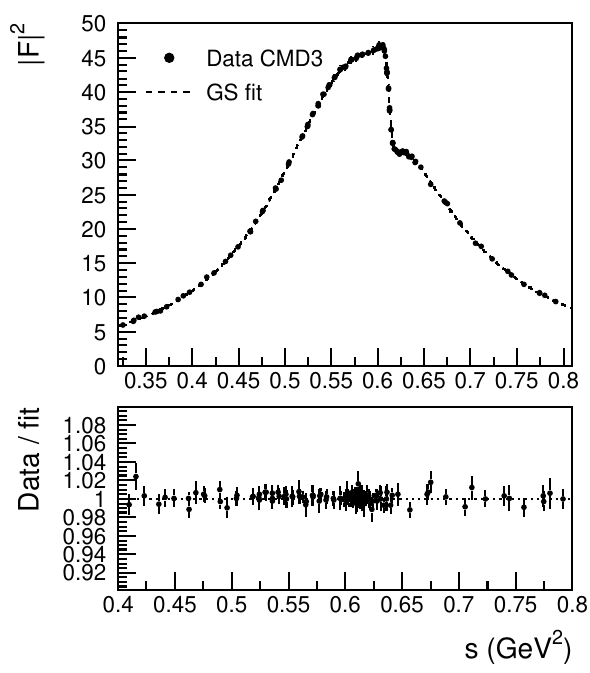}
    \includegraphics[width=0.42\columnwidth]{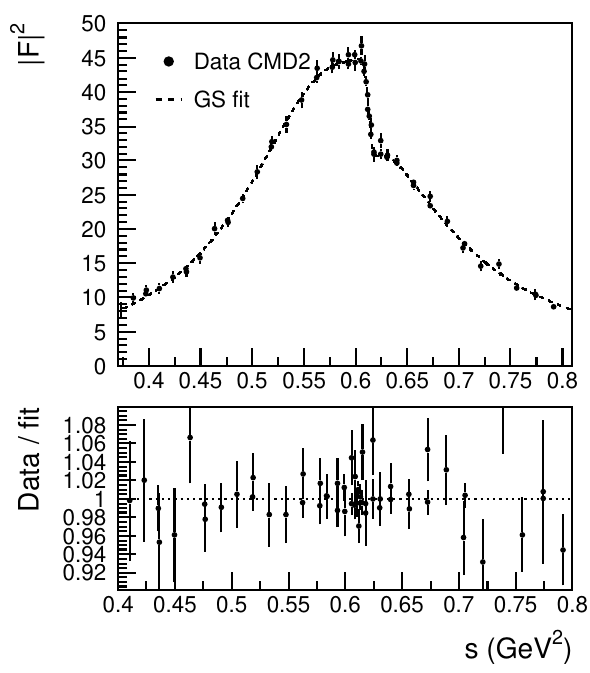}
    \includegraphics[width=0.42\columnwidth]{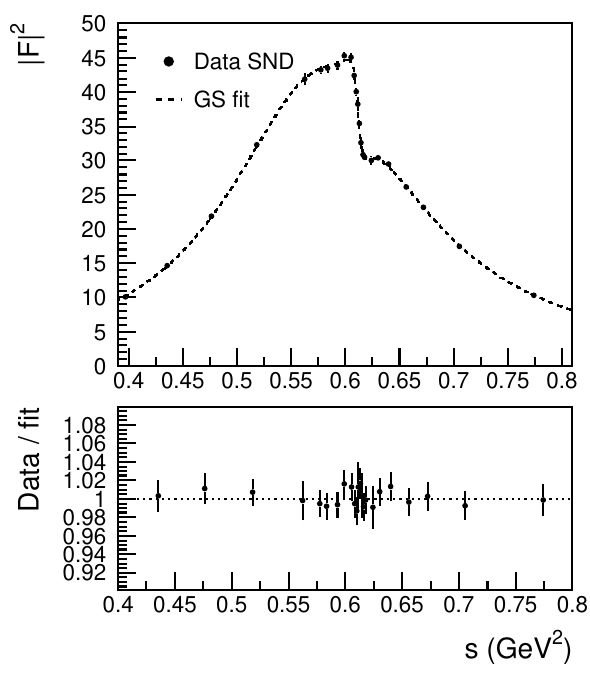}
    \includegraphics[width=0.42\columnwidth]{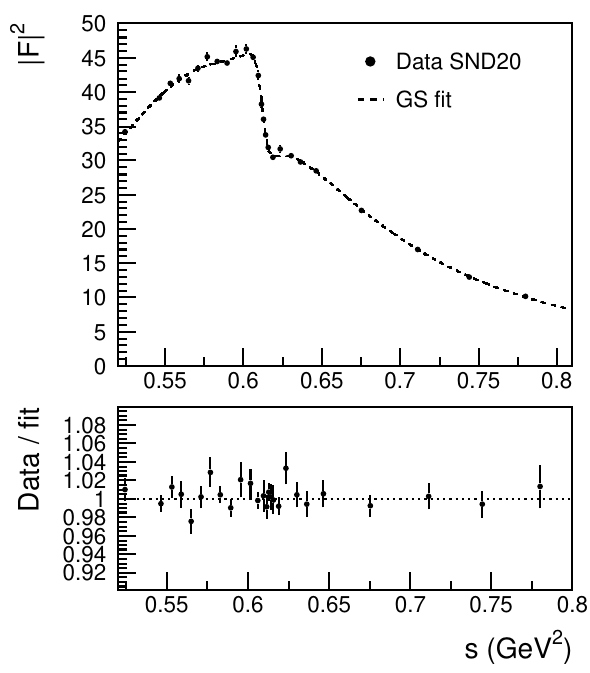}
    \includegraphics[width=0.42\columnwidth]{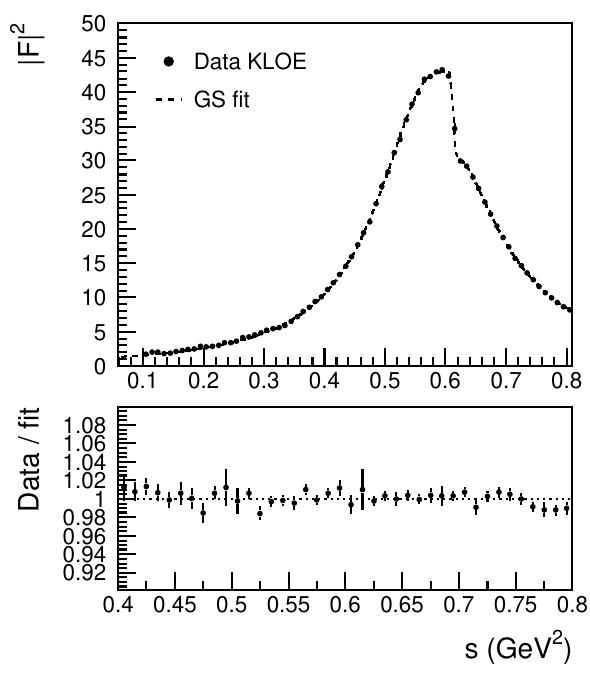}
    \includegraphics[width=0.42\columnwidth]{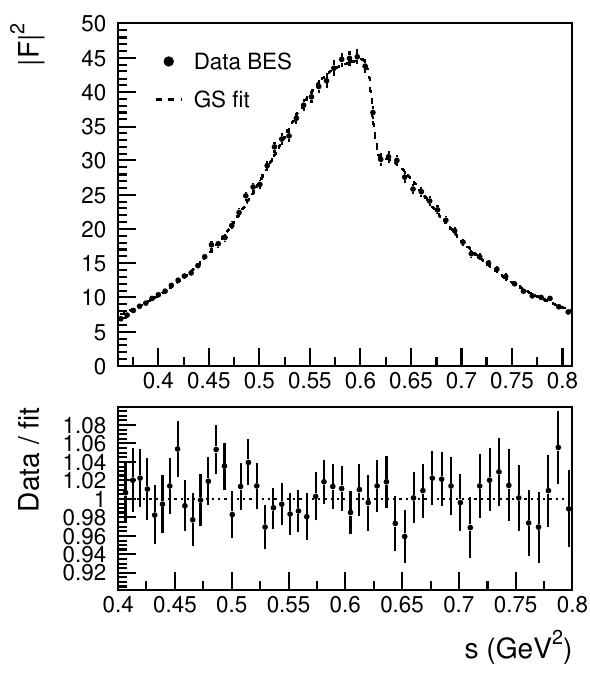}
    \caption{\label{fig:eefits}   \small Fits to $e^+e^-$ data with GS $\rho$ resonance function and $\rho-\omega$ interference with 7 free parameters as described in the text: from top-left to bottom-right CMD-3, CMD-2, SND, SND20, KLOE, BES. For each experiment, the top panel shows the pion form factor decoupled from normalization (data points) and the fitted function, with their ratio in the bottom panel. The ratio for BABAR is given in Figure~\ref{fig:tail} (top-left).
}
\end{figure}

\begin{figure}[tbp] \centering
    \includegraphics[width=0.42\columnwidth]{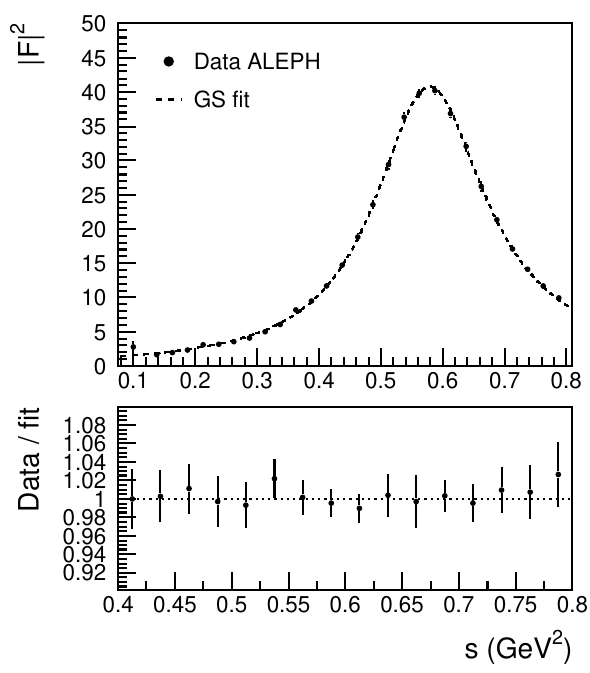}
    \includegraphics[width=0.42\columnwidth]{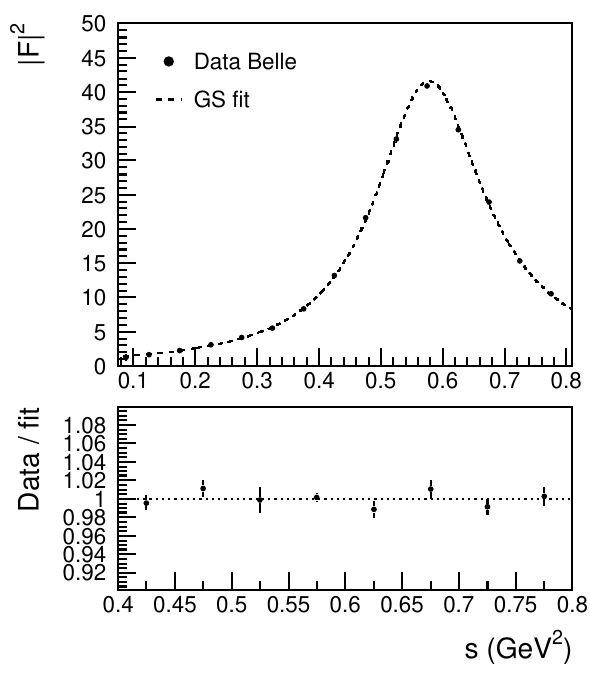}
    \includegraphics[width=0.42\columnwidth]{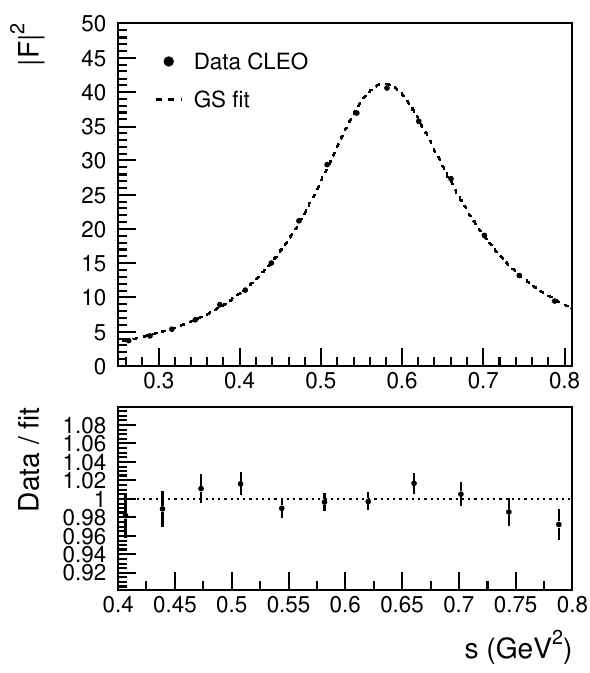}
    \caption{\label{fig:taufits}   \small Fits to $\tau$ data with GS $\rho$ resonance function with three free parameters as described in the text: ALEPH~(top-left), Belle~(top-right) and CLEO~(bottom). For each experiment, the top panel shows the pion form factor decoupled from normalisation (data points) and the fitted function, with their ratio in the bottom panel.
}
\end{figure}

The full-range BABAR fit is displayed in Figure~\ref{fig:tail}, showing the fast-falling resonance tail with well-reproduced structures up to 3\,GeV. The consistency of the BABAR tail with CMD-3 and Belle data is also shown. Fits of the other experiments are given in Figure~\ref{fig:eefits} for $e^+e^-$ and Figure~\ref{fig:taufits} for the $\tau$ data. Ratios between the data and the fitted function are also presented to check the quality of the fits. The results for the fitted parameters are found in Table~\ref{tab:ee-fit} for $e^+e^-$ and in Table~\ref{tab:tau-fit} for $\tau$. All the $\chi^2$ values are good, except for the Belle fit. In that case we had to make choices in order to match the information on uncertainties provided in the publication with additional material made available to us~\cite{Hayashii} for our past work on combined $\tau$ spectral functions~\cite{Davier:2013sfa}. Among these choices, the most conservative estimate of uncertainties has been employed here.

\begin{table}[tbp]
\centering
\label{tab:ee}
\caption{\small Results of the GS fit to $e^+e^-$ data. The numbers in the parentheses correspond to the fitted uncertainties. The $\chi^2$ value of the fit per number of degrees of freedom (DF) is also shown.}\label{tab:ee-fit}
\vspace{2mm}
\resizebox{\textwidth}{!}{
\begin{tabular}{|l|c|c|c|c|c|c|c|} \hline
 Parameter               & BABAR      & BES          & CMD2         & CMD3       & KLOE       & SND          & SND20 \\\hline
Normalization            & 1.031(14)  & 0.990(15)    & 0.996(12)    & 1.052(8)   & 1.004(6)  & 1.013(16)    & 1.020(15) \\
$m_\rho$ (MeV)           & 775.21(33) & 775.63(75)   & 775.70(48)   & 774.92(9)  & 774.44(18) & 775.26(39)   & 775.57(32) \\
$\Gamma_\rho$ (MeV)      & 149.45(67) & 146.56(2.00) & 145.98(1.38) & 148.74(21) & 149.05(48) & 148.47(1.20) & 148.41(1.25)\\
$m_\omega$ (MeV)         & 782.02(15) & 782.56(1.23) & 782.66(28)   & 782.32(6)  & 783.28(41) & 782.25(27)   & 782.21(16)\\
$\Gamma_\omega$ (MeV)    & 8.05(24)   & 10.2(2.7)    & 8.7(5)       & 8.45(9)    & 9.7(7)     & 8.50(35)     & 8.40(38)\\
$|c_\omega|$ (10$^{-3}$) & 1.64(5)    & 1.89(41)     & 1.63(12)     & 1.73(2)    & 1.55(7)    & 1.69(8)      & 1.72(8)\\
$\phi_\omega$ ($^\circ$) & 1.5(2.3)   & 11(12)       & 9.6(3.9)     & 4.5(7)     & 4.9(2.1)   & 8.5(3.3)     & 6.2(2.4)\\\hline
$m_{\rho^\prime}$ (GeV) & 1.464(16) & -- & -- & -- & -- & -- & --\\
$\Gamma_{\rho^\prime}$ (GeV) & 0.285(22) & -- & -- & -- & -- & -- & --\\
$|c_{\rho^\prime}|$ & 0.15(1) & -- & -- & -- & -- & -- & -- \\
$\phi_{\rho^\prime}$ ($^\circ$) & 202(6) & -- & -- & -- & -- & -- & -- \\
$m_{\rho^{\prime\prime}}$ (GeV) & 1.825(18) & -- & -- & -- & -- & -- & -- \\
$\Gamma_{\rho^{\prime\prime}}$ (GeV) &  0.285(22) & -- & -- & -- & -- & -- & -- \\
$|c_{\rho^{\prime\prime}}|$ & 0.063(6) & -- & -- & -- & -- & -- & -- \\
$\phi_{\rho^{\prime\prime}}$ ($^\circ$) & 46(13) & -- & -- & -- & -- & -- & -- \\
$m_{\rho^{\prime\prime\prime}}$ (GeV) & 2.232(14) & -- & -- & -- & -- & -- & -- \\
$\Gamma_{\rho^{\prime\prime\prime}}$ (GeV) & 0.060(64) & --  & -- & -- & -- & -- & -- \\
$|c_{\rho^{\prime\prime\prime}}|$ (10$^{-3}$) & 3.3(1.6) & -- & -- & -- & -- & -- & -- \\
$\phi_{\rho^{\prime\prime\prime}}$ ($^\circ$) & $-9(30)$ & -- & -- & -- & -- & -- & -- \\\hline
$\chi^2/\mathrm{DF}$ & 0.99 & 0.87 & 0.85 & 0.92 & 1.2 & 0.85 & 1.5 \\
 \hline
\end{tabular}
}
\end{table}
\begin{table}[btp]
\centering
\caption{\small Results of the GS fit to $\tau$ data. The numbers in the parentheses correspond to the fitted uncertainties. For Belle and CLEO, additional calibration and resolution systematic uncertainties on $m_\rho$ and $\Gamma_\rho$ from their publications are also included. The $\chi^2$ value of the fit per number of degrees of freedom (DF) is also shown.}\label{tab:tau-fit}
\vspace{2mm}
\begin{tabular}{|l|c|c|c|} \hline
 Parameter          & ALEPH        & Belle        & CLEO         \\\hline
Normalization       & 1.012(17)    & 1.002(10)    & 1.023(10)    \\
$m_\rho$ (MeV)      & 776.63(1.10) & 775.04(59)   & 775.60(0.99) \\
$\Gamma_\rho$ (MeV) & 151.62(2.17) & 147.45(1.85) & 150.04(1.39) \\\hline
$\chi^2/\mathrm{DF}$ & 0.64 & 1.8 & 1.2  \\
 \hline
\end{tabular}
\end{table}

\section{Consistency checks}

Tests are performed in order to check the validity of some of the assumptions made in this work. The following issues are investigated: (1) The decoupling between resonance parameters and normalization needs to be tested. (2) The question of the possible dependence on a specific parametrization of the $\rho$ resonance line shape is also raised. (3) The sensitivity of the results on the $\rho$ resonance parameters to the description of the high-mass tail of the form factor above 0.9\,GeV is studied, especially since the BABAR results have been used throughout. (4) The effect of the model-dependent $G_\mathrm{EM}(s)$ is investigated. The corresponding checks are presented in turn.

\subsection{Decoupling from normalization}
\label{sec:norm}

The goal of the method used here is to decouple the resonance parameters from the overall absolute scale by fitting a global normalization constant. This is necessary in order for the form factor corrections to be independent from the respective magnitude of the $e^+e^-$ and $\tau$ data. However, this separation is not unique and will likely depend on the chosen parametrization for the form factor. Also we know that serious discrepancies exist between the different $e^+e^-$ datasets~\cite{Davier:2023fpl} which are still not understood. Therefore it is important to check the validity of our procedure and guarantee that the respective $\rho$ parameters can be safely extracted to allow a meaningful comparison between $e^+e^-$ and $\tau$. Since the discrepancies among $e^+e^-$ data are maximal in the $\rho$ resonance region this check is all the more important.

In practice, experimental effects in the different measurements leading to the observed discrepancies in the $\rho$ region may affect both the normalization and the shape. At the level of precision currently achieved, one can characterize these deviations in first-order by differences in normalization. From the detailed study of the cross-section measurements, some trends are seen beyond normalization differences. A second-order description would involve a slope throughout
the rho region. This is indeed observed to be the case between KLOE and BABAR, and even more pronounced between KLOE and CMD-3, the two most distant measurements. By only factorizing the normalization difference in our approach, such an effect will generate a shift in the fitted $\rho$ mass. It is indeed what we observe between KLOE and the other experiments in Table~\ref{tab:ee}. This effect leads to an enlargement of the uncertainty which is taken into account in the average $\rho$ mass determination. Third-order effects affecting the width determination could be present as well, but they are beyond the statistical power of the present data. This is confirmed by the stability of the width values in Table~\ref{tab:ee} between experiments, in particular the consistency between KLOE and CMD-3.

A final test is performed to check the decoupling between the $\rho$ resonance parameters determined in our procedure from the absolute cross-section differences between experiments. This is important in order to verify the independence of the $\tau$/$e^+e^-$ form factor corrections and the derived $a_\mu^\tau$ from $a_\mu^{ee}$. To this end, we compare the $a_\mu^{ee}$ values found in the $\rho$ region from each experiment to the corresponding normalization factors in our fits. In order to accommodate all experiments in this test, the range of integration for the dispersion integral is restricted to 0.6--0.883\,GeV (upper limit fixed by SND20, only slightly reduced compared to the 0.9\,GeV used in all fits). It is seen in Figure~\ref{fig:norm-amu} that the normalization found in the fits tracks very well with the corresponding $a_\mu^\mathrm{HVP,\, LO}$ values. The result is striking for the two most precise, and most different, KLOE and CMD-3 results: whereas their $a_\mu$ values in this range differ by 5.1$\sigma$, the ratios $a_\mu$/normalization are consistent at 0.4$\sigma$ level. The same behavior is observed in the comparison between KLOE and BABAR with values of 2.9$\sigma$ and 0.02$\sigma$, respectively, and for all experiments as shown in Figure~\ref{fig:norm-amu}. This test validates our procedure of decoupling the absolute normalization from the measurement of the form factor parameters. Our results also show that, once normalization differences are taken into account, all $\rho$ parameters are consistent between all 7 experiments, with the exception of the mass value fitted for KLOE, which is the consequence of an additional slope with respect to the other experiments.

\begin{figure}[htp] \centering
    \includegraphics[width=0.695\columnwidth]{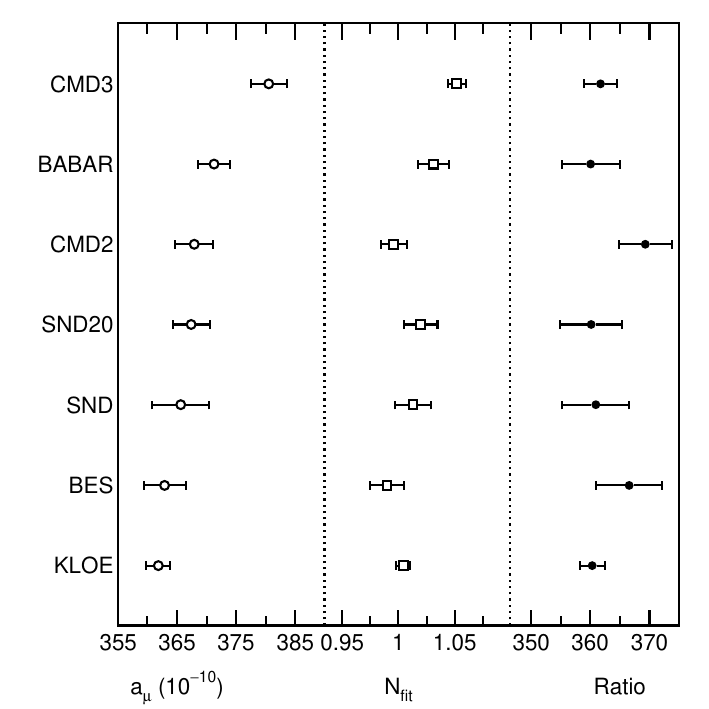}
    \caption{\label{fig:norm-amu}   \small \label{fig:norm-amu}Comparison for all $e^+e^-$ experiments of their $a_\mu^\mathrm{HVP,\,LO}$ integral in the mass range 0.6--0.883\,GeV (left) and the normalization factor in the corresponding GS fits, $N_\mathrm{fit}$ (middle). To better track the correlation, the experiments are ranked by the values of $a_\mu^\mathrm{HVP,\, LO}$ they provide. The two quantities are seen to track each other well, as emphasized in their ratio plot (right) which is remarkably constant. A quantitative comparison is given in the text for the two most precise and most inconsistent experiments, KLOE and CMD-3, and also for KLOE and BABAR.
}
\end{figure}

\subsection{Line shape parametrization}

Our method relies on the actual form factor measurements from $e^+e^-$ and $\tau$ decay spectral functions. The goal is to evaluate their potential differences, originating from IB, so that corrections can be applied to the $\tau$ form factor in order to predict an independent $e^+e^-$-like form factor. The practical issue is to characterize these differences with the smallest set of parameters consistent with the experimental accuracy of the actual measurements. For this, a good parametrization of the data is needed. 

Several examples of $\rho$ resonance parameterizations are available in the literature. The most economic one is the Guerrero-Pich~\cite{Guerrero:1997ku} function inspired by Chiral Perturbation Theory which depends on only one parameter, the $\rho$ mass, the width being predicted by theory. While being elegant, this functional form does not provide good fits to the data at the level of precision required~\cite{Castro:2024prg}. Two-parameter representations are the standard Breit-Wigner shape with mass-dependent width referred to as Kuhn-Santamaria~\cite{Kuhn:1990ad} (KS)~\footnote{The KS parametrization is equivalent to the GS expressions given in Eq.~(\ref{eq:GS}) by setting to 0 the functions $d(m)$ and $f(s,m,\Gamma)$.} and the more elaborate GS form~\cite{Gounaris:1968mw} suited to a wide resonance and with the advantage of satisfying analyticity constraints.  Here, we provide a test using for the most precise $e^+e^-$ data from CMD-3, which dominate the average and the combined $\tau$ data.

\begin{figure}[tbp] \centering
    \includegraphics[width=0.65\columnwidth]{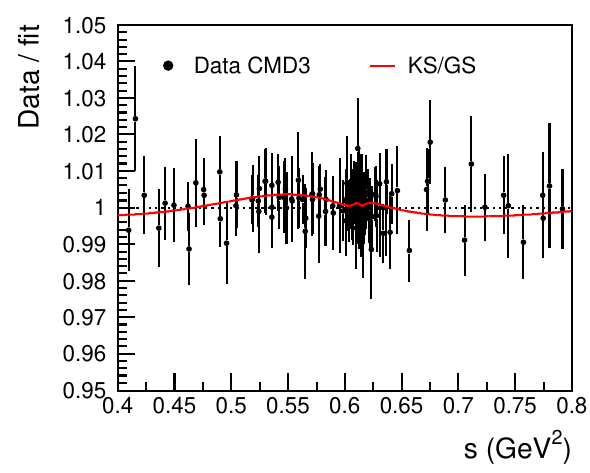}
    \caption{\label{fig:KS-GS}   \small The ratio between data (points) for CMD-3 and the GS fit (dashed line at 1) compared to the result (red curve) fitted on the same data obtained using the KS parametrization. The difference is much smaller than the data uncertainty. Both fits are good representations of the CMD-3 data.
}
\end{figure}
 
Both GS and KS fits provide a satisfactory description for the measured form factor.
The $\chi^2$ for the KS fit is a bit worse ($\chi^2/\mathrm{DF}=1.09$) than for GS ($\chi^2/\mathrm{DF}=0.91$), but still good. The KS fitted result for CMD-3 is shown in Figure~\ref{fig:KS-GS} relative to the GS fit. The two results are slightly different but well within the accuracy of the present data. Therefore, the two parameterizations are suitable to represent the most accurate measurements. However, the derived mass and width values are different in the two cases: for CMD-3 GS and KS yield mass values $(774.92\pm0.09)$\,MeV and $(773.99\pm0.13)$\,MeV, respectively, and for the width $(148.74\pm0.021)$\,MeV and $(147.30\pm0.26)$\,MeV. Clearly no particular physical significance should be attached to these values as they are just parameters of a given parametrization describing well in a phenomenological way the measured $\rho$ line shape. Similar shifts are observed in the $\tau$ fits. The differences between the $\tau$ and $e^+e^-$ values for the two fits are:
\begin{eqnarray}
\Delta m_\rho^\mathrm{GS} &\!\!\!=\!\!\!& -0.30\pm0.38~~{\rm MeV} \\
\Delta m_\rho^\mathrm{KS} &\!\!\!=\!\!\!& -0.91\pm0.48~~{\rm MeV} \\
\Delta \Gamma_\rho^\mathrm{GS} &\!\!\!=\!\!\!& -0.58\pm1.01~~{\rm MeV} \\
\Delta \Gamma_\rho^\mathrm{KS} &\!\!\!=\!\!\!& -0.83\pm1.00~~{\rm MeV}\,.
\end{eqnarray}

However, the parametrization has to be used again when computing the effect of the mass and width $e^+e^-/\tau$ differences on the dispersion integrals. When this is done, the final impact of the KS-GS models on the $a_\mu^\mathrm{HVP,\, LO}$ corrections is found to be:
\begin{eqnarray}
\Delta a_\mu^{m_\rho~\mathrm{GS-KS}} &\!\!\!=\!\!\!& 0.02\times10^{-10} \\
\Delta a_\mu^{\Gamma_\rho~\mathrm{GS-KS}} &\!\!\!=\!\!\!& 1.39\times10^{-10}
\end{eqnarray}
for a total effect on $a_\mu^\mathrm{HVP,\, LO}$ from the pion form factor
between GS and KS of $1.39\times10^{-10}$. This is taken as a systematic uncertainty on the form factor parametrization which is included in the final result shown in the next sections.

\subsection{Sensitivity to the high-mass tail} 

\begin{figure}[htp] \centering
    \includegraphics[width=0.695\columnwidth]{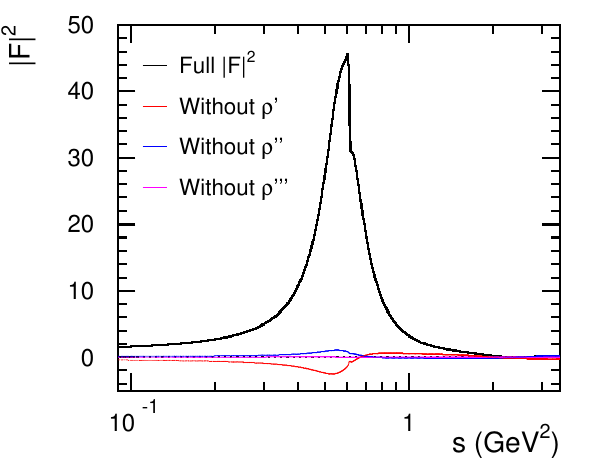}
    \caption{\label{fig:tails-rho}   \small The contribution of the higher $\rho$-like states under the $\rho$ resonance from the central parameters of the BABAR fit. The effect is dominated by the $\rho-\rho'$ and $\rho-\rho^{\prime\prime}$ interferences.
}
\end{figure}

Figure~\ref{fig:tails-rho} shows the contribution of the higher $\rho$-like states under the $\rho$ resonance from the central parameters of the BABAR fit. The dominant effect is from the $\rho-\rho'$ interference affecting mostly the determination of the $\rho$ mass. Any difference between the $\rho'$ parameters between $e^+e^-$ and $\tau$ would produce a bias. In this context, the most sensitive parameter is the $\rho'$ amplitude and the bias would depend on the amount of IB at this level, but specific tests are necessary.

The consistency of the high-mass BABAR tail can be compared to existing $e^+e^-$ data. The precision CMD-3 results extend to 1.2\,GeV and the comparison to the fit performed up to 0.9\,GeV with the BABAR tail is displayed in Figure~\ref{fig:tail}. The comparison is limited to a small mass range, showing maximal deviations at the level of 5\%. Convoluting with the pattern seen in Figure~\ref{fig:tails-rho} the resulting effect is very small. 

The reliability of the BABAR tail as applied to $\tau$ data is studied in the same spirit with the Belle spectrum as also shown in Figure~\ref{fig:tail}. The agreement is overall satisfactory in the full mass range on average, although in detail some oscillations reaching 5\% are observed in the lowest range. Thus the conclusion is similar as for CMD-3.

Another way to test this compatibility is to perform fits of $\tau$ data in the full mass spectrum by fixing the mass and width of all high-mass resonances while keeping free the complex amplitudes for the $\rho'$ and $\rho^{\prime\prime}$ resonances. A direct check of the tail consistency and its impact on the fitted $\rho$ parameters is achieved by comparing the values obtained for these parameters with those of the standard fit with the full BABAR tail. The found values with the free amplitudes are $m_\rho=774.85$\,MeV compared to $775.23$\,MeV in the standard fit. Corresponding values for $\Gamma_\rho$ are $149.57$\,MeV and $149.32$\,MeV. The observed differences of 0.38\,MeV for the mass and $-0.25$\,MeV for the width are taken as respective systematic uncertainties and added to the $\tau$ uncertainties.

\subsection{$G_\mathrm{EM}(s)$ correction}

A possible bias from the model-dependent $G_\mathrm{EM}(s)$ correction is investigated next. First, the two published calculations using Chiral Perturbation Theory~\cite{Cirigliano:2002pv} and a Vector Dominance model~\cite{Flores-Baez:2007vnd} are in agreement. Second, the global effect of the full correction (only the shape in the $\rho$ region is relevant here) is found with the Belle data to have a significant effect on the fitted $\rho$ parameters, although smaller than the corresponding uncertainties. Comparing the two available calculations, the effect on the fitted $\rho$ mass and width is smaller than 0.1\,MeV, thus negligible.

\section{Discussion of the results}

\begin{figure}[tbp] \centering
    \includegraphics[width=0.495\columnwidth]{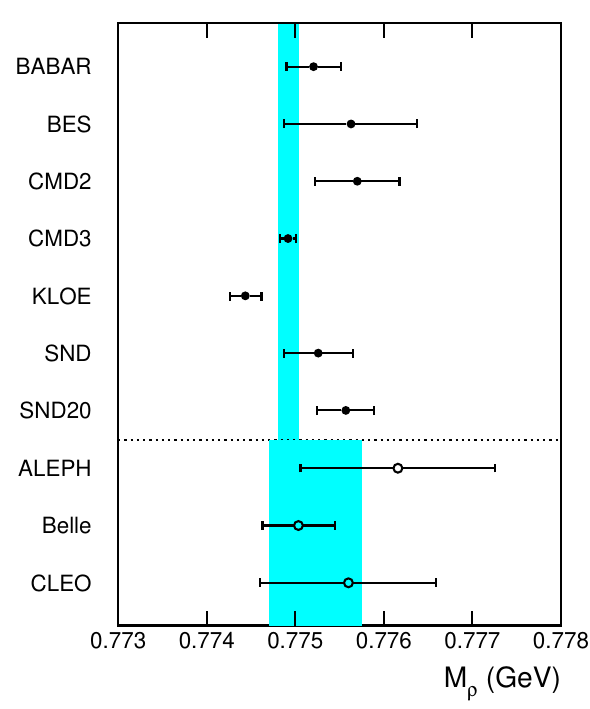}
    \includegraphics[width=0.495\columnwidth]{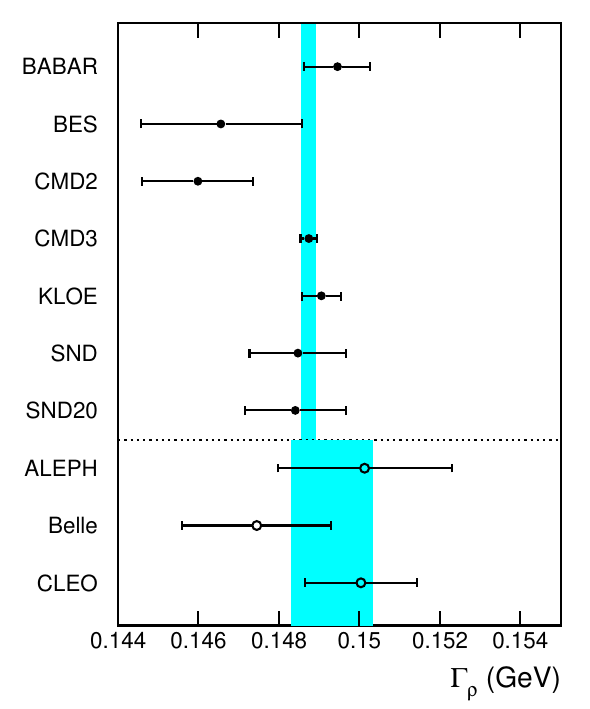}
    \caption{\label{fig:rho-ee-tau}   \small Left: A compilation of the fitted $\rho$ mass values from $e^+e^-$ data in the top panel in comparison with the fitted values from $\tau$ data. Right: the same plot compilation for the $\rho$ width. The vertical bands correspond to the weighted averages of the $e^+e^-$ and $\tau$ determinations, respectively. In the former case, they are scaled by $\sqrt{\chi^2/\mathrm{DF}}$.
}
\end{figure}

In this work, we have presented results from a systematic study of the shape of the respective $e^+e^-$ and $\tau$ two-pion spectral functions under the same conditions. They allow one to compare in a model-independent way the mass and width obtained from the common Gounaris-Sakurai resonance parametrization, independently from the spectral function normalization and thus decoupled from the respective HVP contributions from $e^+e^-$ and $\tau$ input data. The obtained results for all experiments are summarized in Figure~\ref{fig:rho-ee-tau} (left) for the mass and (right) for the width.  The $e^+e^-$ values are dominated by the CMD-3 measurement.

The values for the mass and width differences, $\Delta m_\rho$ and $\Delta\Gamma_\rho$, are obtained from a weighted average of the results from all $e^+e^-$ and $\tau$ experiments which have been shown to be consistent, with the exception of the $\rho^0$ mass value from KLOE
(see the discussion in Section~\ref{sec:norm} for an explanation of this effect). The uncertainties on the weighted averages are scaled by $\sqrt{\chi^2/\mathrm{DF}}$ when it exceeds unity. This scaling only occurs for the mass determination in $e^+e^-$ data with a factor equal to 1.7. The average values are:
\begin{eqnarray}
m_{\rho^0}&\!\!\!=\!\!\!&(774.93\pm0.12) ~~ {\rm MeV} \\
\Gamma_{\rho^0}&\!\!\!=\!\!\!&(148.74\pm 0.19) ~~ {\rm MeV} \\
m_{\rho^\pm}&\!\!\!=\!\!\!&(775.23\pm0.36\pm0.38~[0.52]) ~~ {\rm MeV} \\
\Gamma_{\rho^\pm}&\!\!\!=\!\!\!&(149.32\pm 0.99\pm 0.25~[1.02]) ~~ {\rm MeV}
\end{eqnarray}
where the second uncertainty for the charged $\rho$ is from the high-mass tail and the combined uncertainty is given between square brackets, leading to the differences:
\begin{eqnarray}
\Delta m_\rho&\!\!\!=\!\!\!&m_{\rho^0} - m_{\rho^\pm}=(-0.30\pm0.53) ~~{\rm MeV} \\
\Delta \Gamma_\rho&\!\!\!=\!\!\!&\Gamma_{\rho^0} - \Gamma_{\rho^\pm}=(-0.58\pm1.04) ~~{\rm MeV}\,.
\end{eqnarray}
These values can be compared with the PDG compilation~\cite{ParticleDataGroup:2024cfk}
of published $\rho^0$ and $\rho^\pm$ values, yielding $\Delta m_\rho^\mathrm{PDG}= (0.15\pm0.41)$\,MeV and $\Delta\Gamma_\rho=-(1.7\pm1.1)$\,MeV. Some differences with the present results are seen, but it should be remarked that published individual values rely on fits with different parameterizations, mass ranges, treatments of high-mass tail, not using a floating normalization, and not taking into account  the specific $G_\mathrm{EM}(s)$ radiative correction.

Our results can also be compared with the values obtained in a completely different situation by KLOE~\cite{KLOE:2003kas} from a Dalitz plot fit of the decay $\phi\rightarrow\rho\pi$ involving both charged and neutral $\rho$ mesons. This analysis yielded $\Delta m_\rho^\mathrm{PDG}=+(0.4\pm0.9)$\,MeV and $\Delta\Gamma_\rho=+(3.6\pm2.5)$\,MeV.
The mass difference is consistent with our result, but the width difference is not compatible, however three times less precise. Also, it is noted that the values found in this KLOE analysis for the $\rho$ masses and widths are significantly larger and smaller, respectively, compared to the PDG values. A potential problem is the line shape distortion imposed by the small available phase space for di-pion masses above the $\rho$ peak.

Recent determinations~\cite{Davier:2023fpl,Castro:2024prg} of the two-pion contribution to the LO HVP $a_\mu$ prediction using $\tau$ data used IB form factor corrections based on theory~\cite{Flores-Baez:2006yiq}. For the $\rho$ mass difference, a value $-(1.0\pm0.9)$\,MeV was used in good agreement (0.7$\sigma$) with our data-based result, which is a factor of 2 more precise. The theoretical prediction  for the width difference was $+(0.76\pm0.18)$~MeV, also in agreement (1.3$\sigma$) with the data-based value, but the latter is less precise by a factor of 6. Indeed, the precision obtained from data does not allow checking the calculation of the width difference at the level of accuracy claimed in the prediction.   

\section{Correction for $\rho-\omega$ interference}

\begin{figure}[tbp] \centering
    \includegraphics[width=0.65\columnwidth]{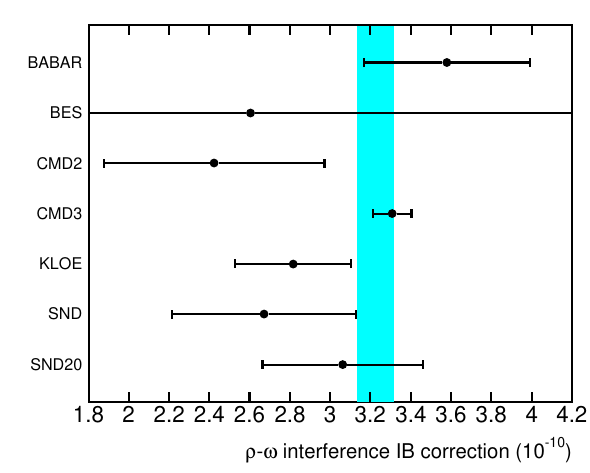}
    \caption{\label{fig:interf}   \small The values for IB correction from $\rho-\omega$ interference, $\Delta a_\mu^{\mathrm{HVP,\, LO},\, \rho-\omega}$, obtained from the $e^+e^-$ fits to all experiments. The weighted average is indicated by the vertical band.
}
\end{figure}

As a by-product of our systematic $e^+e^-$ fits, parameters for $\rho-\omega$ interference, another IB correction to be applied to $\tau$ data, are obtained. The result for the corresponding $\Delta a_\mu^\mathrm{HVP,\, LO}$ is then obtained for each experiment. The individual values are shown in Figure~\ref{fig:interf}. Good agreement between the results is observed for which a weighted average is computed, yielding 
$\Delta a_\mu^{\mathrm{HVP,\, LO},\, \rho-\omega} = +(3.22\pm0.09)\times 10^{-10}$, dominated by CMD-3.
This value is used in the following section for the re-evaluation of the complete IB corrections.

\section{Impact on the $a_\mu$ HVP determination using $\tau$ two-pion data}

We are now in a position to ascertain the impact of our data-based determination of the IB corrections to the pion form factor for the HVP contribution to $a_\mu$ from $\tau$ two-pion data. All contributions are re-evaluated taking advantage of the better parametrization of the spectral functions. In practice, the values are very close to our previous analysis~\cite{Davier:2023fpl}, with the exception of the $\rho-\omega$ interference and mostly the pion form factor correction as the result of the present data-based study. The individual components are listed in Table~\ref{tab:damu}.

\begin{table}[tb]
\centering
\caption{\small The different contributions for IB corrections $\Delta a_\mu^\mathrm{HVP,\, LO}$ applied to the $\tau$ result (in units of $10^{-10}$). For each entry, statistical and systematic uncertainties are combined and given in the first bracket, while the second bracket in the last three entries indicates the uncertainty from the form factor parametrization.}\label{tab:damu}
\vspace{2mm}
\begin{tabular}{|l|c|} \hline
Source of IB    &  $\Delta a_\mu^\mathrm{HVP,\, LO}$~$(10^{-10})$ \\\hline
$S_\mathrm{EW}$       & $-12.21(15)$    \\
$G_\mathrm{EM}$      & $-1.87(96)$  \\ 
FSR    &  +4.65(44) \\
$\beta^3$ cross-section  &$-7.69$   \\
$\rho-\omega$ interference   &  +3.22(9)(8)  \\
Form factor $\Delta m_\rho$   &  +0.06(10)(2)   \\
Form factor $\Delta \Gamma_\rho$  &  +1.62(2.92)(1.39)  \\\hline
Sum      & $-12.22(3.41)$  \\
 \hline
\end{tabular}
\end{table}

 Summing all the contributions, the total IB correction to $a_\mu$ comes out to be:
\begin{equation}
\Delta a_\mu^\mathrm{HVP,\, LO~\tau~IB} = -(12.2\pm 3.4)\times 10^{-10}\,.
\end{equation}

In our last determination~\cite{Davier:2023fpl}, the total IB correction based on the theoretical estimate for the pion form factor was found to be $-(14.9\pm 1.9)\times10^{-10}$. The more recent work in Ref.~\cite{Castro:2024prg} using the same theoretical approach for the IB correction provided a value $-(15.2^{+2.3} _{-2.6})\times10^{-10}$, in agreement with our previous determination. The data-based correction found in the present analysis is consistent with these two results, with however a smaller value and an increased uncertainty. It removes the issue related to the reliability of the calculation of radiative corrections, affecting the charged/neutral $\rho$ mesons by substituting the theoretical estimate with a determination based only on data.

\section{Conclusion and prospects}

\begin{figure}[tb] \centering
    \includegraphics[width=0.65\columnwidth]{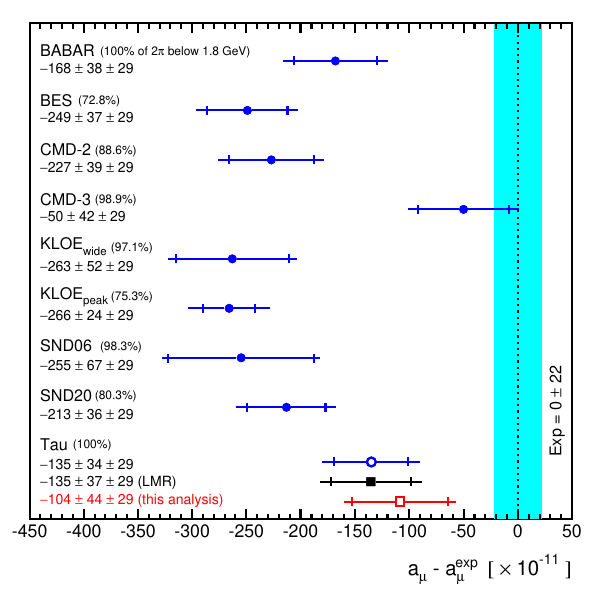}
    \caption{\label{fig:amu-comp}   \small The landscape of dispersive data predictions to $a_\mu$ using $e^+e^-$ et IB-corrected $\tau$ data expressed as differences to the direct Fermilab measurement~\cite{Muong-2:2023cdq}(adapted from Ref.~\cite{Davier:2023fpl}). The full circles correspond to $e^+e^-$ experiments, the open circle to the previous $\tau$ value~\cite{Davier:2023fpl}, the closed square to the $\tau$ estimate from LMR in Ref.~\cite{Castro:2024prg}, both using theory-based IB corrections, and the open square is the result presented in this analysis using data-based corrections. For all points, the inner error bar represents the total uncertainty from the two-pion contribution (the first quoted numerical uncertainty) and the outer one the total uncertainty when adding in quadrature the other contributions to the full $a_\mu$ (corresponding to the common uncertainty of 29 units). The value given in brackets indicates the fraction of the $a_\mu^{2\pi}$ contribution which is covered by the given experiment, the rest being complemented by the combination of all other experimental inputs. 
}
\end{figure}

The alternative of using $\tau$ data rather than $e^+e^-$ is conditioned on being able to compute in a reliable way the isospin breaking occurring between the corresponding spectral functions. In the dominant two-pion channel, IB corrections have so far been determined using theory. The largest uncertainty arises from IB in the pion form factor usually parametrized by the charged/neutral $\rho$ masses and widths. In this letter, we studied the possibility to replace the theory-driven approach by a new method based only on experimental data. In order to retain the independence between $\tau$ and $e^+e^-$ input to the HVP dispersion integral, the normalization of the respective spectral functions is decoupled from the shape of the form factors parametrized using the Gounaris-Sakurai resonance function. Fits of all available precise data have been performed in similar conditions in a systematic way, yielding an excellent description of all the datasets. The obtained mass and width differences are then used to compute the corrections to be applied to $\tau$ data. Many consistency checks have been performed to test the reliability of the procedure and estimate the corresponding systematic uncertainties. The re-evaluation of all IB components leads to a value of $-(12.2\pm 3.4)\times 10^{-10}$ for the total IB correction.

The impact of this new value can be appreciated by incorporating it in the full $a_\mu$ prediction complementing the 2$\pi$ contribution by non-2$\pi$ hadronic components from $e^+e^-$ compilation and other non-HVP-LO contributions as done in Ref.~\cite{Davier:2023fpl} with information therein. The obtained $a_\mu$ can then be compared to all other predictions from $e^+e^-$ experiments, with their known lack of consistency. Compared to our previous result~\cite{Davier:2023fpl}, the new value is slightly displaced upward, close to the CMD-3 ($0.9\sigma$) and BABAR ($1.1\sigma$) results, but further away from KLOE ($3.3\sigma$ or $2.4\sigma$ depending on the KLOE range considered). Figure~\ref{fig:amu-comp} displays all the results expressed as differences with the current Fermilab direct measurement~\cite{Muong-2:2023cdq}. The $\tau$ result deviates from the direct measurement by $1.8\sigma$.

This data-based method is limited by the precision of the available $\tau$ $2\pi$ data. Therefore, improvement is expected with the forthcoming analysis of a much larger dataset being produced at the Belle\,II experiment. Another interesting possibility exists using the huge sample of $J/\psi$ decays recorded by the BES\,III experiment, amounting to about $10^8$ $J/\psi\rightarrow \rho \pi$
decays in which all $\rho$ charge states can be studied in a very clean environment. Therefore, prospects for improvement look good.
\printbibliography
\end{document}